\documentclass[namedreferences]{SolarPhysics}
\usepackage[optionalrh,natbib]{spr-sola-addons} 
\usepackage{graphicx}                    
\usepackage{amssymb}                     
\usepackage{color}                       
\usepackage{url}                         

\newcommand{\degree}{^{\circ}}

\newcommand{\aap}{    {\it Astron. Astrophys.}}

\newcommand{\apj}{    {\it Astrophys. J.}}
\newcommand{\apjl}{    {\it Astrophys. J. Lett. }}

\newcommand{\grl}{    {\it Geophys. Res. Lett.}}

\newcommand{\jgr}{    {\it J. Geophys. Res.}}

\newcommand{\solphys}{{\it Solar Phys.}}

\newcommand{\ssr}{    {\it Spa. Science Rev.}}
\newcommand{\planss}{    {\it Plan. Spa. Sci.}}

\begin{document}

\begin{article}

\begin{opening}

\title{The Wave-Driver System of the Off-Disk Coronal Wave 17 January 2010}

%
\author{M.~\surname{Temmer}$^{1}$\sep
      B.~\surname{Vrsnak}$^{2}$\sep
        A.M.~\surname{Veronig}$^{1}$}

%
\runningauthor{M.~Temmer et al.}
\runningtitle{Wave-Driver System of an Off-Disk Coronal Wave}

%
  \institute{$^{1}$ Kanzelh\"ohe Observatory/IGAM, Institute of Physics, University of Graz,
Universit\"atsplatz 5, A-8010 Graz, Austria \\  email: \url{manuela.temmer@uni-graz.at} \\
             $^{2}$ Hvar Observatory, Faculty of Geodesy, University of Zagreb,
Ka\v{c}i\'{c}eva 26, HR-10000 Zagreb, Croatia  \\
             }

\begin{abstract}
We study the 17 January 2010 flare--CME--wave event by using STEREO/SECCHI EUVI and COR1 data. The observational study is combined with an analytic model which simulates the evolution of the coronal-wave phenomenon associated with the event. From EUV observations, the wave signature appears to be dome shaped having a component propagating on the solar surface ($\overline{v}\approx$~280~km~s$^{-1}$) as well as off-disk ($\overline{v}\approx$~600~km~s$^{-1}$) away from the Sun. The off-disk dome of the wave consists of two enhancements in intensity, which conjointly develop and can be followed up to white-light coronagraph images. Applying an analytic model, we derive that these intensity variations belong to a wave-driver system with a weakly shocked wave, initially driven by expanding loops, which are indicative of the early evolution phase of the accompanying CME. We obtain the shock standoff distance between wave and driver from observations as well as from model results. The shock standoff distance close to the Sun ($<$0.3~\textit{\textit{R$_\odot$}} above the solar surface) is found to rapidly increase with values of $\approx$0.03\,--\,0.09~\textit{R$_\odot$} which give evidence of an initial lateral (over-)expansion of the CME. The kinematical evolution of the on-disk wave could be modeled using input parameters which require a more impulsive driver ($t$\,=\,90~s, $a$\,=\,1.7~km~s$^{-2}$) compared to the off-disk component ($t$\,=\,340~s, $a$\,=\,1.5~km~s$^{-2}$).
\end{abstract}
%
\end{opening}


\section{Introduction}

In large part, our knowledge of coronal mass ejections (CMEs) comes from coronagraph observations delivering white-light data. CMEs, as observed in white light, often exhibit a typical three-part structure, consisting of a bright rim encircling a dark cavity, mostly followed by a bright core \citep[][]{illing85}. Therefore, by definition, a CME is a structured intensity enhancement observed in white-light. The actual process that launches the ejection is most probably connected to magnetic restructuring. This early evolution phase of a CME can often be observed in the extreme ultraviolet as well as soft X-ray data in the form of expanding loop systems \citep[\textit{e.g.}][]{harrison00,vrsnak04-cme}.

CMEs, as they evolve and propagate away from the Sun, are able to drive magnetohydrodynamical (MHD) shocks in the corona that can be tracked by coronal type II radio bursts \citep{gopalswamy97,magdalenic10}. The formation of the shock itself is dependent on the time--speed profile of the CME as well as on the spatial distribution of the Alfv\'{e}n speed in the solar corona, which in turn is related to the local magnetic-field strength and density of the ambient plasma. To generate a shock, the CME needs to have a sufficiently high velocity with respect to the local Alfv\'{e}n speed and such favorable conditions are assumed to be present in the middle corona over $\approx$2~\textit{R$_\odot$} \citep[\textit{e.g.},][]{gopal01,mann03}. Recent studies showed that shocks driven by fast CMEs are observable in white-light data \citep[\textit{e.g.},][]{vourlidas03,ontiveros09,bemporad10,kim12} as well as EUV \citep[e.g.][]{veronig10,kozarev11,ma11,gopalswamy11,cheng12}, and UV spectra \citep[e.g.][]{raymond00,bemporad10}.

The evolution of a three-dimensional dome connected to a surface shock wave is observed for the 17 January 2010 CME--flare event. It was studied in detail by \cite{veronig10} who showed that the surface as well as the off-limb structure are part of an evolving three-dimensional wave-dome formed by a weak shock. The surface wave propagated with a mean speed of $\approx$280~km~s$^{-1}$ whereas the upward moving part was of much higher speed of $\approx$650~km~s$^{-1}$ \citep{veronig10}. The difference between the speed of the upward-moving part of the wave and the on-disk signature was interpreted by \cite{veronig10} in the following manner: the upward-moving part is driven all of the time by the outward moving CME, whereas the surface signature is only temporarily driven by the flanks of the expanding CME and then propagates freely.

A recent article by \cite{grechnev11}, studying the same event, supports the result that the dome-structure was actually a shock-driven plasma flow. \cite{grechnev11} simulated the evolution of the shock wave from which they concluded that most likely an abrupt eruption of a filament caused the weak shock. They compare this with a blast-wave scenario during which the wave is only briefly driven. \cite{zhao11} investigated, for the 17 January 2010 event the relation between the surface wave speed, the CME speed and the local fast-mode characteristic speed. They concluded that the observed CME front is in fact a wave phenomenon just like the EUV wave on the solar surface.

In this study, we focus on the kinematical evolution of the off-disk signature of the dome-shaped wave event and add new aspects not covered by previous studies. Using observations of the SECCHI instruments EUVI and COR1 on STEREO we will show, by applying an analytical model with input parameters constrained by the observations, that the off-disk signature in fact consists of two components: a driver and a weakly shocked wave. The driver of the off-limb wave evolves from expanding loop structures and is interpreted as the CME, the observed frontal part is interpreted as the shock wave ahead. In particular, we investigate the shock offset (standoff) distance for the wave-driver system.

\section{Data}

The EUVI instrument \citep{wuelser04} and the COR1 and COR2 coronagraphs are part of the \textit{Sun Earth Connection Coronal and Heliospheric Investigation} \citep[SECCHI:][]{howard-stereo08} instrument suite onboard the STEREO mission \citep{kaiser08}, launched in October 2006. On 17 January 2010 STEREO-B/EUVI observed on eastern limb a flare/CME event, associated with a dome-shaped structure which can be observed off-limb as well as on-disk. The upward moving dome is well observed in white-light coronagraph STEREO-B/COR1 and COR2 data. In the following study we use EUVI 171{\AA}~and 195{\AA}~filtergrams with a temporal cadence of 60 and 90~seconds, respectively, as well as COR1 white-light data with a cadence of five minutes. Using these instruments we can follow the event in EUVI out to 1.7~\textit{\textit{R$_\odot$}} and in COR1 over the field-of-view of 1.4 to 4~\textit{\textit{R$_\odot$}}. Accordingly, we focus on the low coronal signatures of the flare/CME event and its early evolution phase. We note that associated with the event is a high-frequency type II burst drifting from $\approx$310 MHz to $\approx$80 MHz during $\approx$03:51\,--\,03:58~UT. The source region of the event under study is active region AR~11041 located at S25E128, \textit{i.e.}\ occulted as viewed from Earth.

In order to derive the kinematical evolution of the off-disk wave, we follow the dome-shaped structure. For this we use i) manual tracking of intensity enhancements and ii) perturbation profiles over the dome structure. The perturbation profiles are defined as intensity variations averaged over angular sectors along the propagation direction of the dome. Since the off-limb wave is not evolving radially from the solar surface \citep[see also][]{grechnev11}, the kinematical profiles are derived along the direction of motion which is $-$12$\degree$ off the radial direction.

\section{Analytical Model}\label{analyt}

\begin{figure}
 \centerline{\includegraphics[width=1\textwidth,clip=]{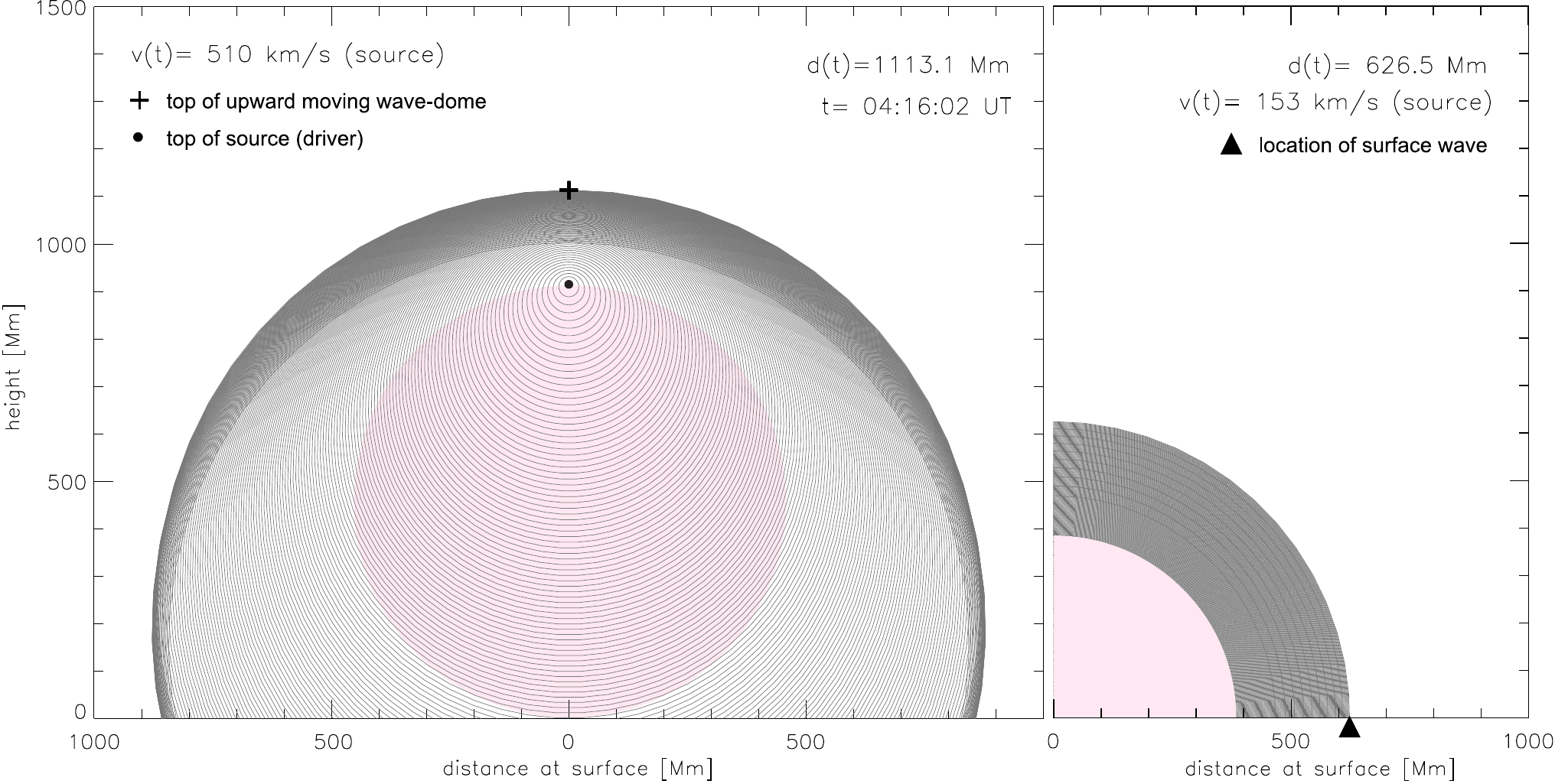}}
 \caption{Snapshot of a model run for 04:16:02 UT. Grey circles indicate the propagation of the wave signals that the outermost part of the driver has sent each ten seconds starting from 03:48:32~UT. The driver velocity [$v$] is given by the distance of the outermost part of the driver at time $t$  with $d(t)$ (\textit{cf.}\ Figure~\ref{model}). The pink shaded circle gives the size of the source (driver). Left: Applying the parameter set which simulates the upward moving dome of the wave, \textit{i.e.}\ the off-disk wave. Right: Applying the parameter set which simulates the on-disk wave. }
    \label{model0}
\end{figure}

Under the assumption that the coronal wave under study is a large-amplitude MHD wave, we show that the observed dome belongs to a wave--driver system. We simulate the observed kinematical profile of the dome structure by using the analytical model developed by \cite{temmer09} which we briefly describe in the following.

In the model, the driver of the wave is a ``synthetic'' source surface which continuously emits MHD signals at time steps of $\Delta t$ = ten seconds. The signals start to be emitted at $t_{0}$ and are iteratively followed at each time step $\Delta t=t_{i}-t_{i-1}$ during its evolution until the time $t_i$. We obtain the distance from the source region center [$d(t_{i})=r(t_0)+x(t_{i})$] where $r(t_{0})$ is the radius of the source surface at the time $t_0$ and $x(t_{i})$ is the distance traveled by the signal from $t_0$ until $t_i$. The geometry of the driver is spherically symmetric and radially expanding with a radius $r(t)$ centered at height $h(t)$. The beginning of the shock formation in the model is determined when a later emitted signal overtakes the outermost one \citep[for more details see][]{vrsnak00a}.

For the driver of the off-disk wave we consider a source that may expand and move at the same time with a constant radius-to-height ratio [$r(t)/h(t)$] acting as a combined bow-shock/piston driver. We refer to a bow-shock for a shock wave that moves with the same speed as the driver and material can flow behind the driver, whereas the piston-driven shock continuously compresses the wave ahead leading to an increase in the shock standoff distance and increase in speed of wave \citep[for more details on the terminology see][]{vrsnak05}. For completeness we also simulate the on-disk surface wave for which we use a synthetic source expanding only in the lateral direction without upward motion, \textit{i.e.}\ plasma cannot flow behind the contact surface and the driver acts as a piston \citep[][]{warmuth07,zic08}. For the one-dimensional case, \cite{vrsnak00a} developed a simplified relation between the propagation speed of the surrounding plasma and the amplitude of the wave. From this it follows that the rest-frame speed of the wave signals [$w$] is related to the flow velocity [$u$] which is associated with the perturbation amplitude, and the local Alfv\'en velocity [$v_{\rm A0}$] as $w=v_{\rm A0}+3u/2$. Since we do not know the spatial distribution of $v_{\rm A0}$ in the corona we simply express the change of $v_{\rm A0}$ with distance using an exponential function \citep[see Equation (2) in][]{temmer09}. In this way, the exponential function regulates the decay of the wave signals. For more details on the model we refer the reader to \cite{temmer09}.

On-disk coronal waves are assumed to be driven impulsively over a short time and then to propagate freely \citep[\textit{e.g.},][]{vrsnak08} whereas, due to the upward movement of the CME, a separate mechanism acts on off-disk waves. Therefore, for simulating the on- and off-disk wave we will apply two different expansion mechanisms of the driver, which enables us to derive their physical characteristics separately.

Figure~\ref{model0} gives a snapshot of a model run showing the wave signals (circles) that were emitted during the expansion of the source. The left panel of Figure~\ref{model0} considers the simulation of the off-disk wave and shows the emitted signals for an upward moving (along the \textit{y}-axis) and simultaneously expanding source. The snapshot presents the time step 04:16:02~UT at which the frontal part of the spherical source has a height of $h(t)$\,=\,914.8~Mm and a radius of $\approx$450~Mm. This is comparable to observational results as given in Figure~\ref{composite}. The right panel of Figure~\ref{model0} shows the emitted signals for a source expanding in the lateral direction (along \textit{x}-axis), simulating the evolving flanks of the CME and the on-disk wave. From the model results we extract the kinematics of the solar surface signal, \textit{i.e.}\ EUV wave, and the kinematics of the summit of the off-disk wave. We stress that the input parameters for the driver as well as the extracted kinematics of the simulated wave are constrained by observational results. Different model runs are performed until a best match is found between the model input/results and the observational results.

\section{Results}

\begin{figure}
\centerline{\includegraphics[width=1\textwidth,clip=]{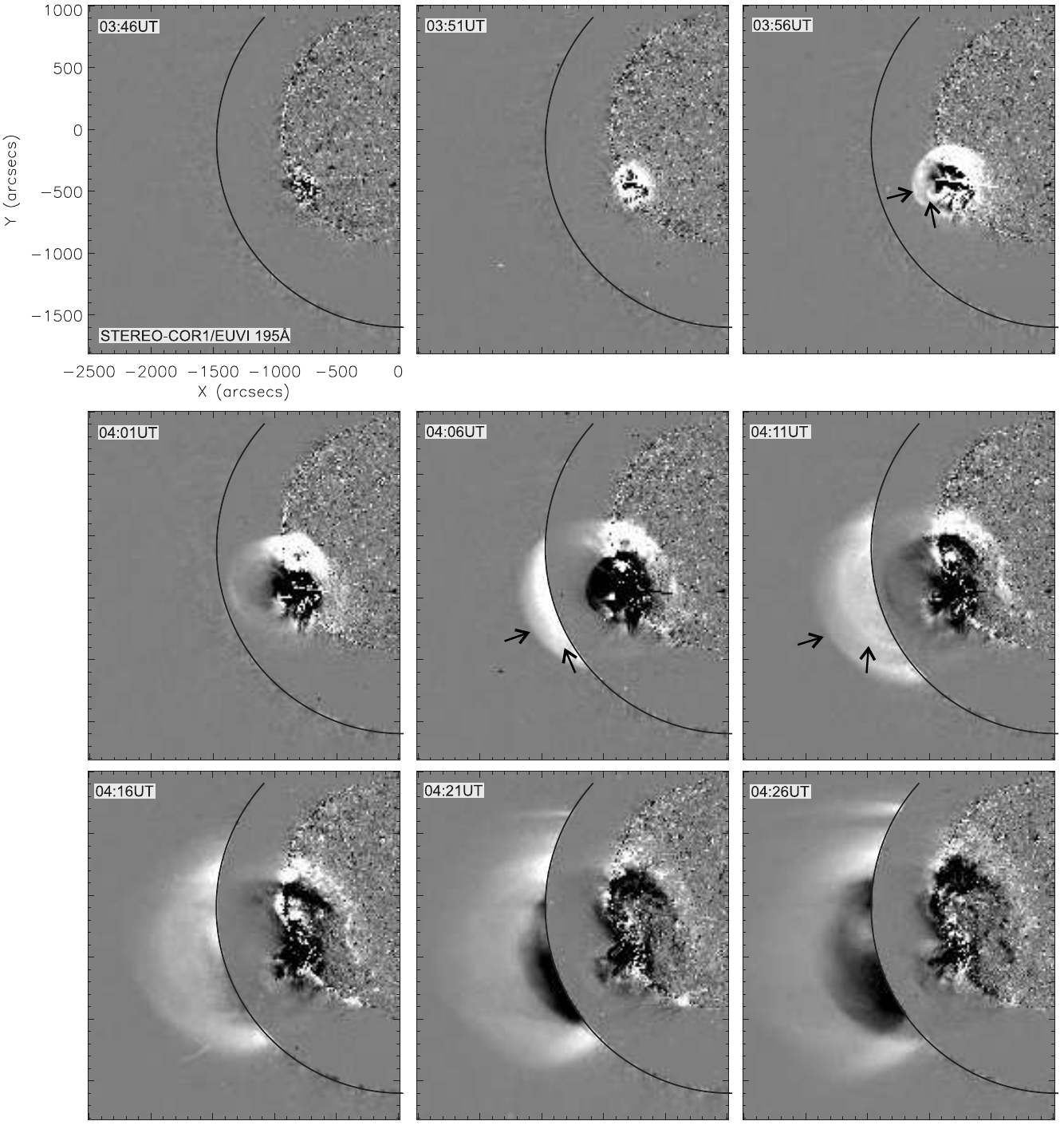}}
 \caption{Composite EUVI 195{\AA}~and COR1 observations from STEREO-B. The running difference images show the evolution of the coronal wave as it propagates off-limb as well as a separate evolving structure below (black arrows). See Electronic Supplementary Material for the accompanied movie.}
    \label{composite}
\end{figure}

Figure~\ref{composite} presents composite images from EUVI~195{\AA}~and COR1, which show the evolution of the surface wave as well as the dome of the wave moving outwards from the Sun in an almost radial direction. The dome of the wave can be seamlessly connected to the surface wave which supports the idea that it is part of the on-disk wave and not the frontal part of the erupting structure \citep[see also][]{veronig10}. In addition, a secondary intensity enhancement is observed behind the top part of the wave, which can be interpreted as the driver of the wave. The top part of the wave and the structure behind evolve and expand concurrently since no black/white feature is visible in the running-difference image. This gives support that the top part of the wave and the secondary intensity enhancement belong together and form a wave--driver system. We note that the two components can not be resolved beyond 2~\textit{R$_\odot$} which restricts our analysis to the early evolution phase of the wave-driver system.

\begin{figure}
\centerline{\includegraphics[width=1\textwidth,clip=]{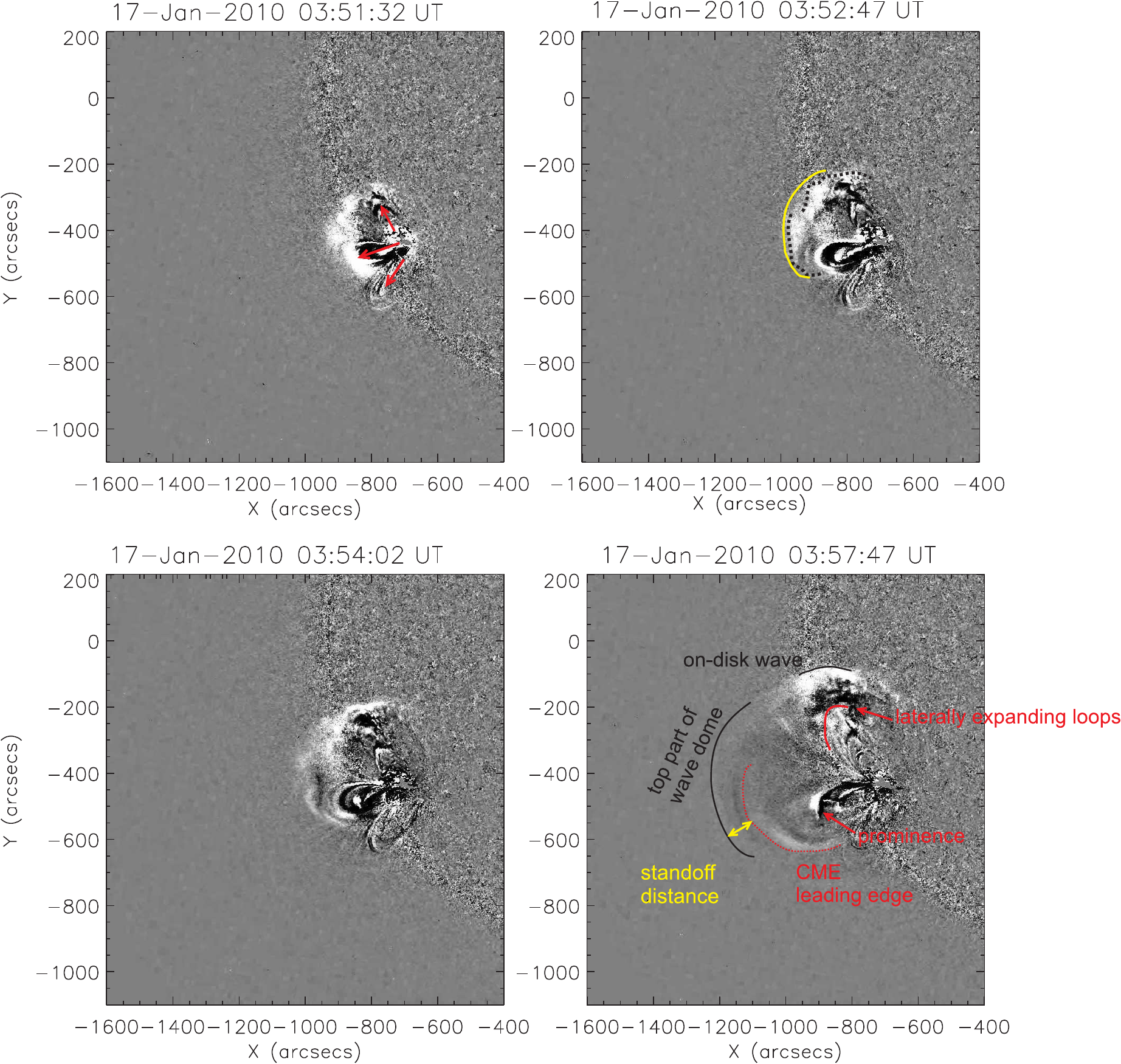}}
 \caption{EUVI-B 195{\AA}~running-difference images showing in detail the evolution of different loop systems and features (labeled) associated with the event. }
    \label{detail}
\end{figure}

Figure~\ref{detail} gives details about the active region as observed with EUVI-B~195{{\AA}} showing the evolution of different loop systems. In total we identify three loop systems (marked with red arrows in the top left panel) evolving in different directions from the active region. One to the northern direction, one radially away from the Sun and one off-disk directed to the South. We note that the northern loop structure observed at 03:51:32~UT may look similar to a wave signature but may actually be a loop system pushed down as a consequence of the lateral expansion of the central eruption \citep{patsourakos10}. At 03:52:47~UT a circularly shaped signature, presenting the coronal wave, appears (marked with a yellow line). In addition, some internal structure visible as intensity enhancements (black dashed line) are observed behind the wave front and ahead of the expanding loops, most probably resulting due to compressed plasma. The on-disk wave evolves from the northern loop system and the off-disk signature of the wave becomes visible at $\approx$0.28~\textit{R$_\odot$} above the solar surface ahead of the radially expanding loop structure. From the southern loop system less distinct coronal wave signatures evolve. We note that the surface coronal-wave could be observed with highest intensity in the northward direction \citep{veronig10}. During the early evolution of the off-disk wave, a clear spatial gap between the two intensity enhancements is observed which is increasing with time (see also Figure~\ref{composite}). We interpret the first intensity enhancement as the top part of the wave dome and the secondary intensity enhancement as the leading edge of the driver, \textit{i.e.}\ CME, the spatial gap between as shock-standoff distance. This gives further evidence for a wave--driver system.

\begin{figure}
\centerline{\includegraphics[width=1\textwidth,clip=]{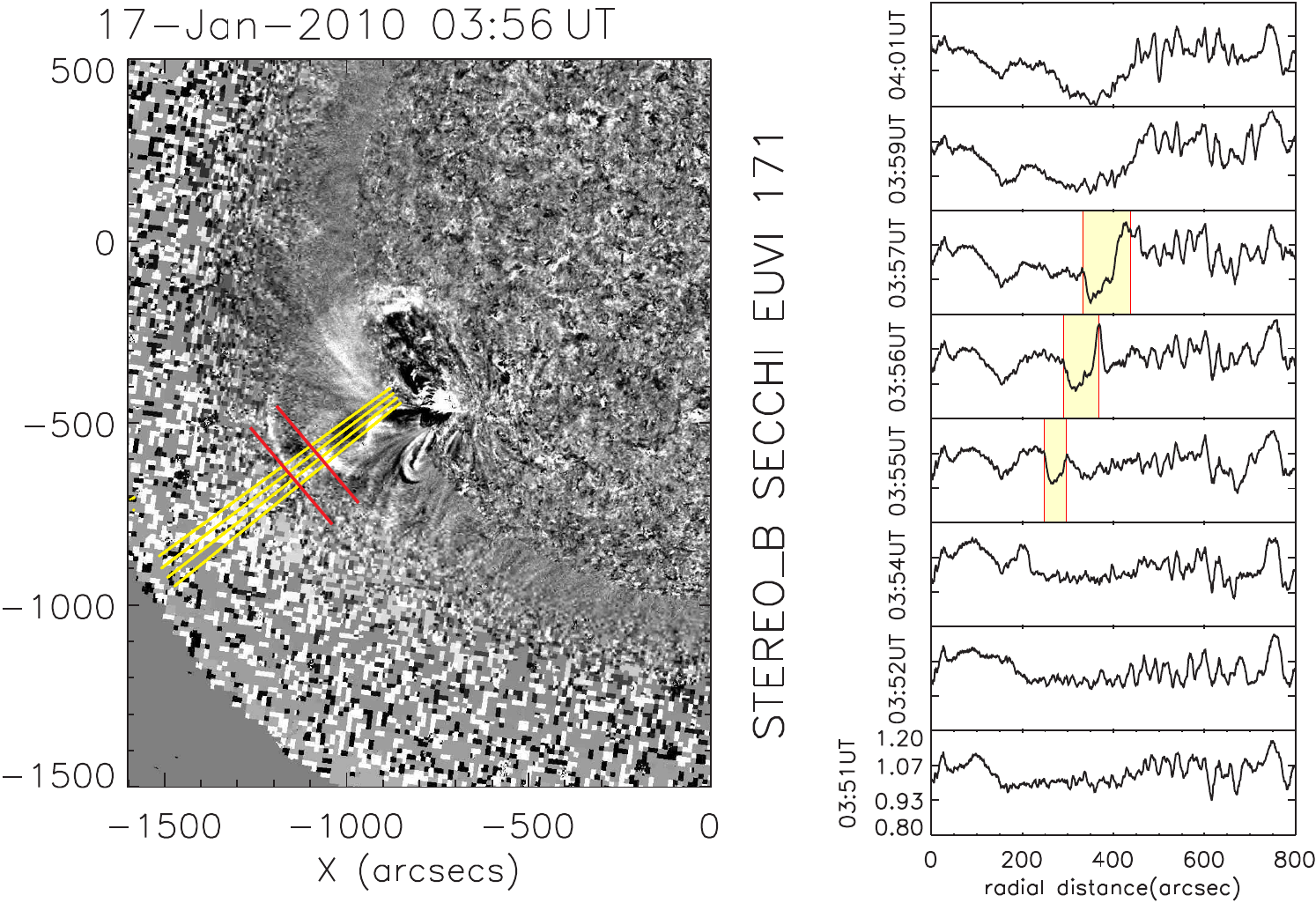}}
 \caption{Left: EUVI-B 171\,{\AA}~ratio image (03:56~UT\,$-$\,03:40~UT) and intensity profiles of the off-limb structure measured along its propagation direction. Yellow lines mark a region of 5$\degree$ over which the mean brightness is calculated, red lines mark the identified driver and wave component. Right: Profiles derived for different time steps showing changes of intensity relative to a pre\,-\,event image given as function of distance above the solar surface. The standoff distance is marked with a yellow shaded area, red lines mark the identified driver and wave component. }
    \label{ratio}
\end{figure}

From base-ratio EUVI-B~171{\AA}~images (Figure~\ref{ratio}), we derive profiles of changes in the intensity relative to a  pre-event image (03:40~UT). The intensity profiles are calculated by averaging the intensity variation of the image over an angle of 5$\degree$ above the solar surface along their direction of motion. The left panel of Figure~\ref{ratio} shows a ratio image together with the region (yellow lines) over which the mean brightness is obtained. The derived averaged intensity is given as function of distance above the solar surface in the right panel of Figure~\ref{ratio}. This clearly shows a spatial gap in the brightness of the dome-shaped wave structure, which can be followed during three time steps. At 03:55~UT the wave and driver components show up first in the profile, having similar relative intensity enhancement of $\approx$7\% above background level. The profiles of time steps 03:56~UT and 03:57~UT reveal that, relative to the driver component, the wave gains in intensity and the distance between them increases. This can be interpreted as compression of plasma ahead of the driver and steepening of the wave front, which propagates faster than the driver.

\begin{figure}
 \centerline{\includegraphics[width=1\textwidth,clip=]{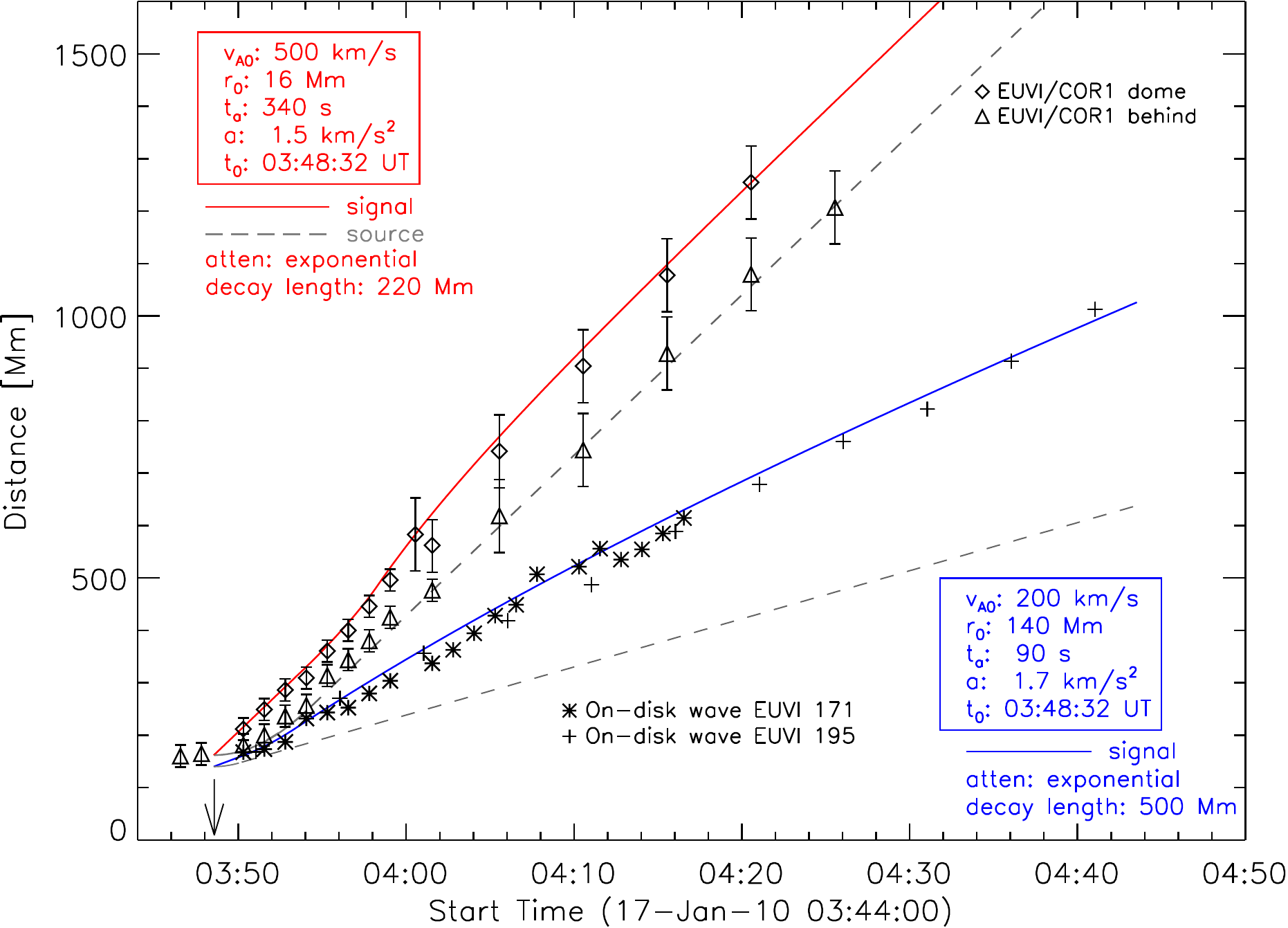}}
 \caption{Distance--time plot for the on-disk wave, the top part of the dome of the wave, and the feature behind the wave. Overplotted is the outcome of model runs, which simulate the propagation of a wave, applying the parameters given in the legend. Red indicates the upward moving dome of the wave, grey dashed its driver, \textit{i.e.}\ the feature behind the wave dome. In blue we present the EUV wave propagating at the solar surface and its driver as a grey-dashed line.}
    \label{model}
\end{figure}

The ratio images become very noisy further out than $\approx$1.3~\textit{R$_\odot$} (\textit{cf.}\ Figure~\ref{ratio}). By using running difference images we derive the distance-time profile of the dome wave over the entire FoV of EUVI-B~195\,{\AA} and 171\,{\AA} and beyond that from COR1 data. We manually measure the top part of the dome as well as the secondary intensity structure along their propagation of motion. The derived kinematics of wave/driver of the off-disk wave is shown in Figure~\ref{model}. The observational results of the solar-surface wave are taken from the study by \cite{veronig10} and are calculated as the mean distance of the wave fronts from the derived wave center along great circles on the solar surface. Furthermore, Figure~\ref{model} presents the simulated kinematics for the off-disk wave and its driver as well as for the on-disk wave, choosing model parameters constrained by observations. The first wave signals are emitted by the synthetic source at $t_{0}$\,=\,03:48:32~UT which is a few minutes before the first wave signature could be identified \citep[see][]{veronig10}. We note that a type II radio burst appeared at 03:51~UT indicating that the wave has to be launched well before \citep[see also][]{grechnev11}.

To simulate the kinematics of the upward-moving off-disk wave together with its driver that best match the observations we use a synthetic driver which accelerates over a time span of 340~seconds with $a$=1.5~km~s$^{-2}$ giving a final velocity of $v$\,=\,510~km~s$^{-1}$. The mean speed of the resulting wave measured $-$12$\degree$ off the radial direction is  $\approx$600~km~s$^{-1}$ \citep[\textit{cf.}][]{grechnev11}. In addition, the source size is set to be proportional to height at each time [$t$] with $r(t)/h(t)$\,=\,0.1. The surrounding Alfv\'{e}n speed of the unperturbed plasma is chosen as $v_{A0}$\,=\,500~km~s$^{-1}$ \citep[see][]{mann03}. This type of source expansion acts as a combined bow-shock/piston driver for the emitted signals. The decay of the signal follows an exponential function and is set, according to the best match between observational and model results, with a decay length of 220~Mm. The distance of the front of the wave minus the top part of the source is defined as the standoff distance. The timing of shock formation is about 03:51:42~UT, which is close to the occurrence of the type II burst, after that the wave is freely propagating.

In order to mimic the surface signal (EUV wave) the source of the surface wave expands in a lateral direction and is fixed at the surface which can be interpreted as piston. The kinematics for this source is a synthetic profile accelerating over a time span of 90~seconds with an acceleration of $a$\,=\,1.7~km~s$^{-2}$ giving a final velocity of the driver of $v$\,=\,153~km~s$^{-1}$. The surrounding Alfv\'{e}n speed of the unperturbed plasma is chosen as $v_{A0}$\,=\,200~km~s$^{-1}$ \citep[see][]{mann03}. The decay length of the wave signals of 500~Mm is chosen to be consistent with observations given in \cite{veronig10}, where the intensity profile strongly decreases in the range of 500\,--\,800~Mm. The timing of shock formation for the solar-surface wave is about 03:50:02~UT after which the wave is freely propagating.

\begin{figure}
\centerline{\includegraphics[width=1\textwidth,clip=]{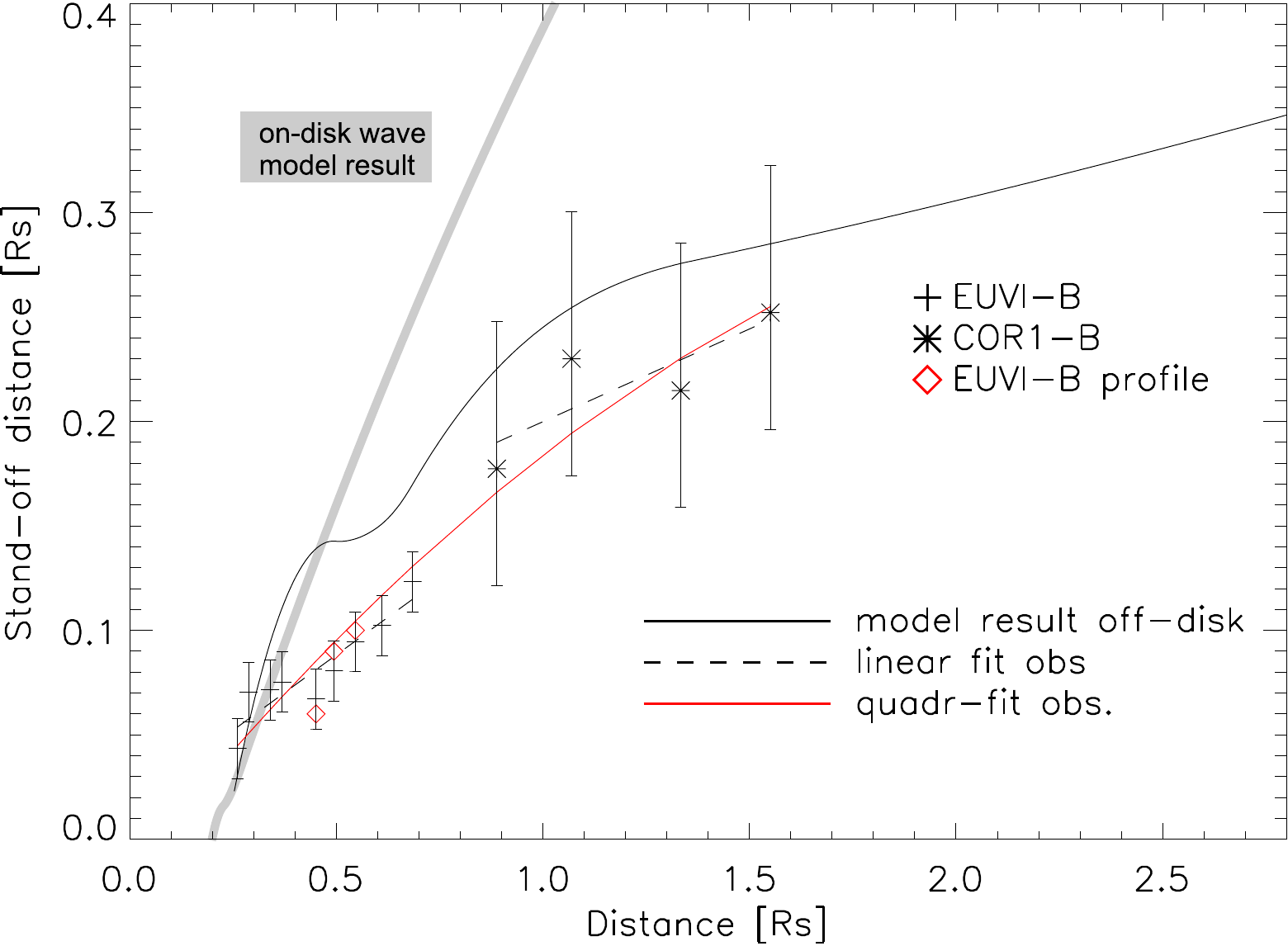}}
 \caption{Standoff distance between the driver and off-disk wave \textit{versus} of propagation distance of the top part of the dome structure (starting from solar surface) derived from EUVI-B and COR1-B observations and the model (solid line). The dashed line gives a linear fit to the observational results separately performed on EUVI-B and COR1-B data, the red line a quadratic fit to all observational results. For completeness we show as a grey line the model results for the standoff distance of the on-disk wave--driver system versus distance of wave front measured from the initiation location. }
    \label{standoff1}
\end{figure}

Figure~\ref{standoff1} shows the shock standoff distance between the driver and wave component for the dome-shaped structure. A good match is found between the standoff distance derived by manually tracking the wave-driver system and extracted from intensity profiles. The observed standoff distance shows a rather linear evolution up to 1~\textit{R$_\odot$}. Beyond this distance, COR1-B observations (last four data points) indicate a decreasing growth-rate of the standoff distance, \textit{i.e.} a certain ``stagnation'' of the growth.

The model results for the off-disk wave, presented by the black solid line in Figure~\ref{standoff1}, show a rather sharp increase of the standoff distance at heights below 0.3~\textit{R$_\odot$}. Beyond this height, a non-linear regime starts, characterized by a stagnation of the growth-rate, quite similar to that found from COR1-B measurements. Such a behavior can be attributed to the way that the source surface, from which the wave signals are emitted, behaves. The source moves upward and expands at the same time with $r(t)/h(t)$, hence the expansion is coupled to the kinematical characteristics of the synthetic source. The standoff distance therefore reflects the kinematical profile of the synthetic source. The stagnation beyond $\approx$1.2~\textit{R$_\odot$} results from the assumed decay of the wave, calculated by an exponential function with a decay length of 220~Mm. This can be interpreted such that the initial lateral (over-)expansion of the CME \citep[][]{patsourakos10} acts as a piston close to the Sun, which results in an increase in the standoff distance since plasma material can not flow behind the driver \citep{warmuth07,zic08}. As the wave--driver system further evolves it becomes more of a piston/bow-shock type and the increase in distance between driver and wave is less strong. The standoff distance derived for the on-disk wave, presented as grey line in Figure~\ref{standoff1}, shows a steep linear increase over distance. This reflects the (3D) piston mechanism of the laterally expanding source which impulsively drives the wave and as the strong expansion of the driver stops the wave continuously separates from the driver.

\begin{figure}
 \centerline{\includegraphics[width=1\textwidth,clip=]{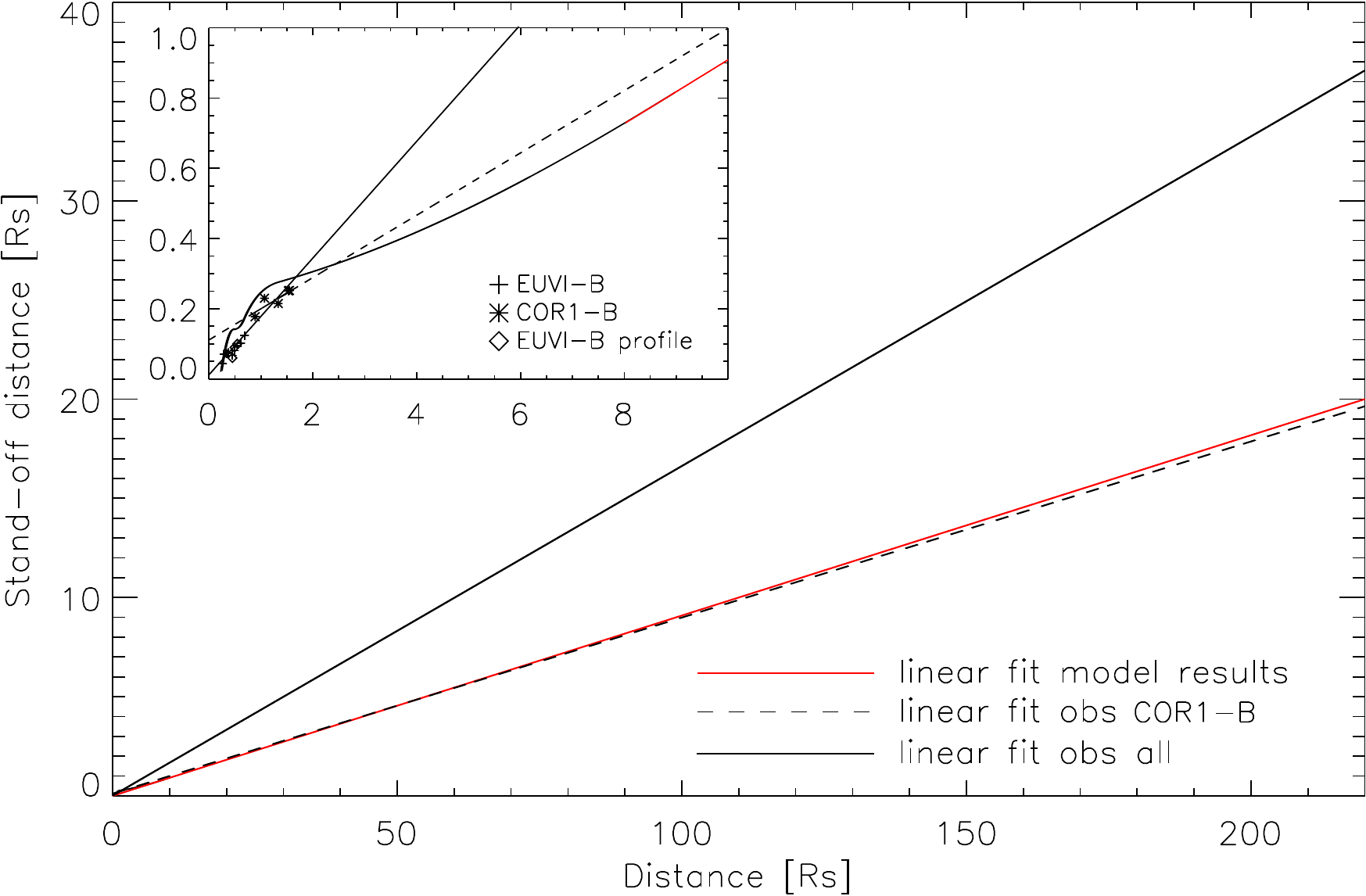}}
 \caption{Standoff distance \textit{versus} distance of top part of dome (starting from solar surface) over the Sun--Earth distance range as derived from linear fits to the model and observational results, respectively. Solid line indicates the linear extrapolation over all observed data points, dashed line is a fit and its extrapolation to COR1-B data points only. The red line shows a linear fit to the model results over the distance range 8\,--\,10~\textit{R$_\odot$}.}
    \label{standoff2}
\end{figure}

Figure~\ref{standoff2} shows an estimation of the shock standoff distance at 1~AU for which we simply extend, by using linear fits (assuming self-similar expansion of the CME), the results derived close to the Sun up to the Earth's location. For the model we use a linear fit to results obtained over the distance range 8\,--\,10\textit{R$_\odot$}. We apply a linear fit to all measured data points and, to take into account the stagnation of the growth rate in standoff distance at larger distance, a linear fit to COR1-B observations only (last four data points). From this we obtain a lower and upper limit for the standoff distance at 1~AU lying in the range of $\approx$20\,--\,36~\textit{R$_\odot$}. We note that the standoff distance derived from the model is very close to the linearly extrapolated standoff distance for COR1-B data points. Considering the wave speed of $v$\,=\,600~km~s$^{-1}$, this corresponds to a shock-CME time lag of the order of 6\,--\,11~hours at 1~AU. This is consistent with reports of the thickness of the magnetosheath of ICMEs measured from in-situ data typically of the order of 0.1~AU \cite[e.g.][]{russell02}. In a recent investigation \cite{maloney11} find a value of 20~\textit{R$_\odot$} for the shock standoff distance at 0.5~AU.

\begin{figure}
\centerline{\includegraphics[width=1\textwidth,clip=]{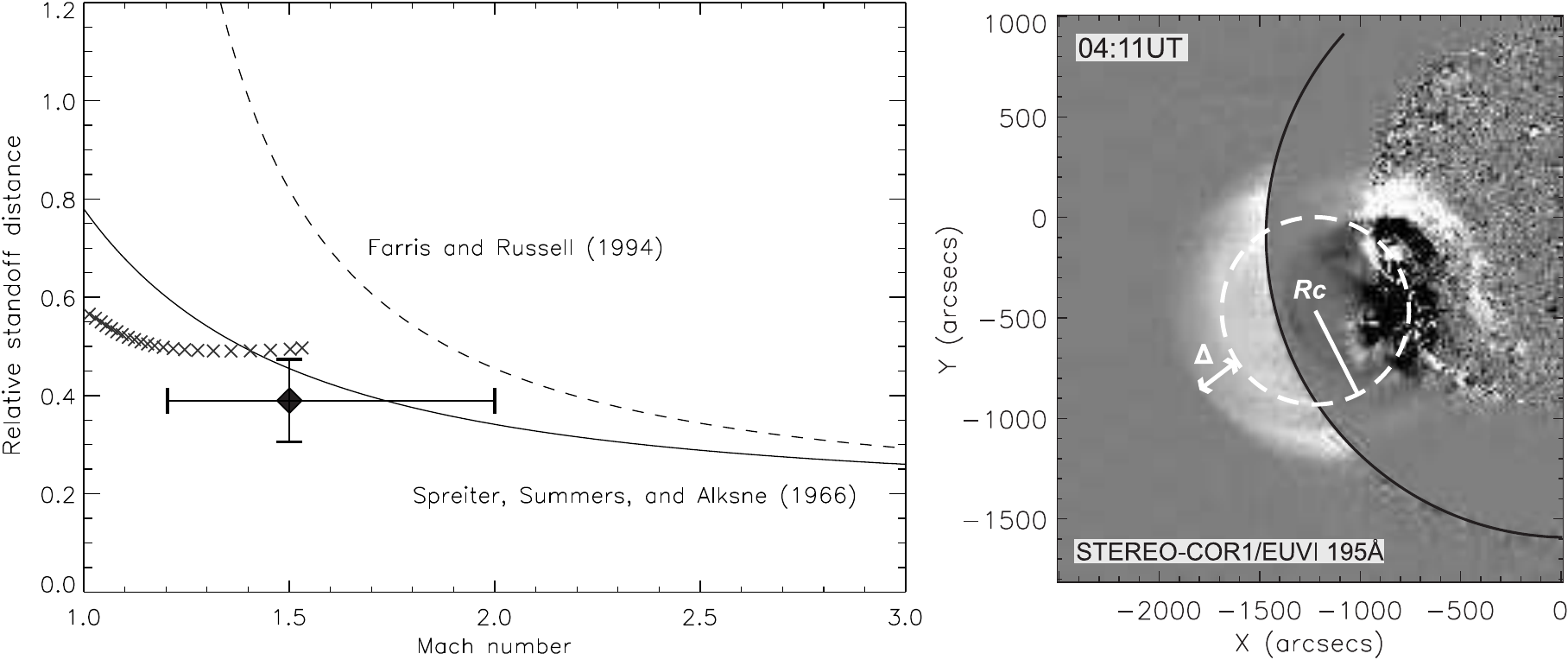}}
 \caption{Standoff distance of a shock from a spherical obstacle with radius of curvature [$R_{c}$] expressed in terms of the radius of curvature for the models of \cite{spreiter66} and of \cite{farris94}. See also Figure~6 in \cite{russell02}. The diamond gives the relative standoff distance [$\Delta/R_{c}$] derived from observations together with error estimations for the uncertainty in the measurements as well as in the unknown value of $v_{A0}$ and consequently $M$. $\times$ represent $\Delta/R_{c}$ derived from the model using $v_{A0}$\,=\,500~km~s$^{-1}$. }
    \label{theory}
\end{figure}

The standoff distance [$\Delta$] between driver and wave is related to the speed and the size of the driver \citep[see, \textit{e.g.},][]{spreiter66,farris94,russell02,zic08}. Physically, it is the Mach number [$M$] and the radius of curvature [$R_{c}$] of the nose of the driver that controls the standoff distance. Therefore, in Figure~\ref{theory} our model results for the relative standoff distance [$\Delta/R_{c}$] are presented as a function of $M$ using $v_{A0}$\,=\,500~km~s$^{-1}$. The model results are compared with one measurement of $\Delta/R_{c}$ determined at a time when the shock structure could be most clearly observed and the kinematical profile of the CME reached constant speed of $\approx$600~km~s$^{-1}$. $R_{c}$ is obtained by fitting a circle to the driver of the wave (see right panel of Figure~\ref{theory}). We derive for the relative standoff distance of the shock under study a value of $\approx$0.4$\,\pm\,$0.1$R_{c}$. Considering a lower and upper limit of the Alfv\'{e}n speed ($300\,<\,v_{A0}\,<\,500$~km~s$^{-1}$) at the measured distance of $\approx$1.5~\textit{R$_\odot$} above the solar surface \citep{mann03}, we obtain $1.2\,<\,M\,<\,2$. Hence, the vertical and horizontal error bars shown on the observational data point reflect the uncertainties in measurements as well as in the derivation of $M$, respectively.

Only as a matter of illustration, in Figure~\ref{theory} we show also the hydrodynamic-model results by \cite{spreiter66} and \cite{farris94} as transformed by \cite{russell02} in terms of $R_{c}$ which read: $$\Delta/R_{c}=0.195+0.585M^{-2}$$ for the high Mach-number approximation by \cite{spreiter66} and $$\Delta/R_{c}=0.195+0.78(M^{2}-1)^{-1},$$ for the version by \cite{farris94} which corrects the previous formula in the low Mach-number regime.

Comparing the data shown in Figure~\ref{theory}, one finds that our model predicts much lower standoff distances at low Mach numbers than both models considered by \cite{russell02}. This is not surprising, since they consider entirely different physical situations. The models by \cite{spreiter66} and \cite{farris94} consider a stationary situation where the obstacle in a supersonic ambient flow has constant size and the ambient flow has steady speed. When applied to CMEs, these models describe a supersonic driver of a constant size and speed, \textit{i.e.}\, a standard bow-shock situation, and at $M\rightarrow 0$ the standoff distance increases $\Delta\rightarrow\infty$. In contrast, our model includes evolutionary aspect, \textit{i.e.}\, considers the acceleration stage of the source surface, which moreover, beside the translatory motion is characterized by expansion, \textit{i.e.}\, it acts not only as a moving object, but also as a three-dimensional piston. The latter effect is especially important since it includes the effect of nonlinear evolution of the wavefront, \textit{i.e.}, its steepening into a shock \citep[see, \textit{e.g.}][]{vrsnak00a}. In such a situation, the time and distance at which the shock forms, and the magnetosheath thickness are determined by the acceleration-time profile of the driver, \textit{i.e.}, its kinematics \citep[see, \textit{e.g.}, Figure~4 in][]{vrsnak00a,zic08}.

Finally, let us note that most observational studies about standoff distances are related to fast CME events and Mach numbers $>$1.5 \citep[see e.g.][]{maloney11,kim12,gopalswamy11}. More observational studies are needed to give a more reliable conclusion about the evolution of the standoff distance for weak shocks.

\section{Discussion and Conclusion}

The 17 January 2010 is a well-observed event revealing a dome-shaped coronal wave structure. In the present study we give evidence from observational and model results that the off-disk part of the wave actually consists of a driver and a wave component. The driver is interpreted as the CME, the frontal part as a weakly shocked wave. We derive that the shock standoff distance shows a linear evolution with a rather rapid increase below 0.3~\textit{R$_\odot$} above the solar surface. These results may be interpreted that the initial lateral (over-)expansion of the CME which is short-lived \citep[$\approx$70~seconds; see][]{patsourakos10} acts as a piston driver to the shock, which leads to a rapid increase in the shock standoff distance. The piston nature of expanding CME flanks is also reflected in results from a recent study by \cite{cheng12}. Using SDO observations, \cite{cheng12} analyzed the structural and kinematical evolution of a CME together with the separation process of a  diffuse wave front from the CME flanks. The wave decoupling from the driver, and with this the actual detection of the wave front, happens when the CME expansion slows down.

Comparing our results to previous studies of off-disk waves moving in the radial direction away from the Sun, we find good agreement for several parameters. We derive for the first observable, \textit{i.e.}, measurable standoff distance values of 0.03\,--\,0.06~\textit{R$_\odot$}. \cite{ma11} who studied a low coronal shock wave using high spatial and temporal resolution data from SDO/AIA report for the thickness of the shocked layer $\approx$0.03~\textit{R$_\odot$}. We find the first signatures of the shock at a distance of $\approx$0.28~\textit{R$_\odot$} above the solar surface which is comparable to the results from \cite{ma11} who find 0.23~\textit{R$_\odot$}. From the model we also derive the timing of the shock signatures to be close to the observed type II radio burst. \cite{cheng12} refer to an almost simultaneous occurrence between a type II radio burst and the start of the separation process between wave and driver. For radio bursts, shock formation heights of $\approx$0.2~\textit{R$_\odot$} are derived by, \textit{e.g.}, \cite{magdalenic10}. We note that the height at which the shock forms is strongly dependent on the speed profile of the driver, \textit{i.e.}\ CME. Peak accelerations of CMEs are found to occur at very low distances from their launch site: $<$0.5~\textit{R$_\odot$} above the solar surface \cite[\textit{e.g.}][]{temmer08,temmer10}.

Using an analytical model, the kinematical profile of both components, driver and wave, can be simulated by applying model parameters that are constrained by observations. In addition, we are able to simulate the on-disk wave using a pure piston-type expansion of the driving source whereas the source of the off-disk wave behaves in the early evolution as piston then becomes more of a bow shock type. We find that the on-disk wave requires a more impulsive driver ($t$\,=\,90~seconds, $a$\,=\,1.7~km~s$^{-2}$) compared to the off-disk wave ($t$\,=\,340~seconds, $a$\,=\,1.5~km~s$^{-2}$). These results lie in between the findings from \cite{grechnev11} who obtain that the off-limb wave was excited impulsively most probably from a filament eruption and then propagated freely, and those of \cite{veronig10} who conclude that the upward-moving dome might have been driven all of the time.

The dome-shaped wave under study evolves from an eruption plus the deformation of different loop systems. In morphology, the dome-shaped wave reminds much more of a CME bubble than of separated loop systems. In visible light, shock waves are reported as well-outlined and sharp boundaries \citep[see][]{ontiveros09}. For the evolution of a coronal surface wave, \cite{temmer10} reported that the wave was launched from two separate centers before it became of circular shape. We may speculate that the loop systems expand and are pushed aside due to the early evolution phase of the erupting structure. The magnetic loop structures form the ``observable envelope'' and are the first signatures of the evolving CME.

The current study shows that relatively slow drivers may cause weak shock waves low in the corona. These waves are visible in white-light and may further propagate up to 1~AU. To investigate the evolution of shock standoff distances in interplanetary space, we require observations of the wave--driver system close to the Sun as well as their in-situ signatures.

 \begin{acks}
MT and AMV greatly acknowledge the Austrian Science Fund (FWF): V195-N16 and P24092-N16. The research leading to these results has received funding from the European Commission's Seventh Framework Programme (FP7/2007-2013) under the grant agreement n$^{\circ}$~263252 [COMESEP] and n$^{\circ}$~284461 [eHEROES].
 \end{acks}

%
%
%
 \bibliographystyle{spr-mp-sola-cnd}

\end{article}
\end{document}